\documentclass[letterpaper,english,aps, twocolumn, prb, longbibliography]{revtex4-1}
\usepackage[T1]{fontenc}

\usepackage{babel}
\usepackage{bm}
\usepackage{amsmath}
\usepackage{amssymb}
\usepackage{graphicx}
\usepackage{wasysym}
\usepackage[unicode=true,pdfusetitle,
 bookmarks=true,bookmarksnumbered=false,bookmarksopen=false,
 breaklinks=false,pdfborder={0 0 1},backref=false,colorlinks=false]
 {hyperref}



\begin{document}
\title{Control of wave scattering for robust coherent transmission in a disordered
medium}
\author{Zhun-Yong Ong}
\email{ongzy@ihpc.a-star.edu.sg}

\affiliation{Institute of High Performance Computing (IHPC), Agency for Science,
Technology and Research (A{*}STAR), 1 Fusionopolis Way, \#16-16 Connexis,
Singapore 138632, Republic of Singapore}
\date{\today}
\begin{abstract}
The spatial structure of the inhomogeneity in a disordered medium
determines how waves scatter and propagate in it. We present a theoretical
model of how the Fourier components of the disorder control wave scattering
in a two-dimensional disordered medium, by analyzing the disordered
Green's function for scalar waves. By selecting a set of Fourier components
with the appropriate wave vectors, we can enhance or suppress wave
scattering to filter out unwanted waves and allow the robust coherent
transmission of waves at specific angles and wavelengths through the
disordered medium. Based on this principle, we propose an approach
for creating selective transparency, band gaps and anisotropy in disordered
media. This approach is validated by direct numerical simulations
of coherent wave transmission over a wide range of incident angles
and frequencies and can be experimentally realized in disordered photonic
crystals. Our approach, which requires neither nontrivial topological
wave properties nor a non-Hermitian medium, creates opportunities
for exploring a broad range of wave phenomena in disordered systems.
\end{abstract}
\maketitle

\section{Introduction}

Understanding how wave scattering in a disordered medium depends on
the spatial structure of its inhomogeneity is important for progress
in fundamental topics such as Anderson localization~\citep{ALagendijk:PT09_Fifty,GModugno:RPP10_Anderson,SGredeskul:LTP12_Anderson,MSegev:NatPhot13_Anderson,AMafi:JLT19_Disordered}
and other transport phenomena~\citep{PSheng:Book06_IntroductionWave}
as well as for a wide range of applications in photonics,~\citep{DWiersma:NatPhotonics13_Disordered,SYu:NatRevMater21_Engineered}
wave shaping,~\citep{HCao:NatPhys22_Shaping} imaging,~\citep{JBertolotti:NatPhys22_Imaging,SGigan:NatPhys22_Imaging}
acoustics~\citep{JMaynard:RMP01_Acoustical}, matter waves.~\citep{JBilly:Nature08_ControlledDisorder}
and even signal filtering.~\citep{FZangenehNejad:AM20_Disorder}
A remarkable insight into this relationship is found in  Ref.~\citep{DKim:PRL22_Bragg}
which shows how the electron wave interaction with the \emph{Berry
potential} can result in sharp Bragg-like scattering in spite of the
absence of periodicity, with the outgoing waves distributed at specific
angles in a manner akin to powder x-ray diffraction. Although this
phenomenon is reported for electron waves in the Schroedinger equation,
its underlying mechanism is not quantum mechanical and instead depends
on the orientation of the incident wave vector $\bm{k}$ with respect
to the Fourier (plane-wave or $\bm{q}$) components of the disorder,\footnote{In our paper, we use the letters $\bm{k}$ and $\bm{q}$ to denote
the wave vectors of the plane-wave states and disorder  Fourier components,
respectively. } as represented by the Berry potential~\citep{DKim:PRL22_Bragg}
\begin{equation}
U(\bm{r};\{\phi_{j}\})=\frac{A}{\sqrt{N}}\sum_{j=1}^{N}\cos(\bm{p}_{j}\cdot\bm{r}+\phi_{j})\ ,\label{eq:BerryPotential}
\end{equation}
where $\bm{r}$ is the position vector in two dimensions, $A$ is
a constant having the dimension of energy, $N$ is the number of Fourier
components,~\footnote{For convenience, we use the term `components' to label the plane-wave
components of the disorder. We reserve the term `state' for the plane-wave
solutions to the Helmholtz wave equation. } and $\bm{p}_{j}$ and $\phi_{j}$ denote the wave vector and phase
of the $j$-th component, respectively, with $|\bm{p}_{j}|=q_{\text{c}}$.~\footnote{Equation~(\ref{eq:BerryPotential}) also includes the contribution
of the component at $-\bm{p}_{j}$ for $j=1,\ldots,N$.} The set of phases $\{\phi_{j}\}$ uniquely determines the spatial
configuration of $U$, with each $\phi_{j}$ taking a value between
$0$ and $2\pi$. Equation~(\ref{eq:BerryPotential}) describes a
two-dimensional (2D) random potential constructed from a superposition
of plane waves of equal wave number $q_{c}$ and distributed over
all propagation angles. Because the Fourier components of the Berry
potential are localized on a circular manifold of radius $q_{\text{c}}$,
electron waves with $|\bm{k}|<q_{\text{c}}/2$ undergo minimal scattering
and thus propagate without attenuation. 

The finding suggests that the distribution of the Fourier components
of the spatial disorder in a medium, as described by its reciprocal-space
($\bm{q}$) spectrum, has profound bearing for scattering and coherent
wave transmission. This principle has been utilized in partially disordered
media,~\citep{SYu:NatRevMater21_Engineered,KVynck:RMP23_Light} in
which the spatial distribution of particulate scatterers, characterized
by the structure factor $S(\bm{q})$, can be correlated to generate
transparency for a range of low-frequency waves.~\citep{OLeseur:Optica16_High}
In \emph{stealthy hyperuniform} (SH) systems in particular,~\citep{STorquato:PhysRep18_Hyperuniform,STorquato:PRE16_Hyperuniformity}
where $S(\bm{q})=0$ for $0<|\bm{q}|\leq q_{\text{SH}}$ and $q_{\text{SH}}$
is the length scale of the absence of long-wavelength density fluctuations,
this results in optical transparency for incident waves with wave
numbers in the range of $|\bm{k}|<q_{\text{SH}}/2$,~\citep{OLeseur:Optica16_High}
a finding closely related to that of Ref.~\citep{DKim:PRL22_Bragg}.
Indeed, one may interpret Eq.~(\ref{eq:BerryPotential}) as an analog
of SH-type disorder for the Schroedinger equation, with the Berry
potential behaving as a random scalar field with stealth hyperuniformity.~\citep{STorquato:PRE16_Hyperuniformity}
Similar conditions for wave transparency in other SH systems have
also been found by Kim and Torquato.~\citep{JKim:NJP20_Effective,STorquato:PRX21_Nonlocal} 

A common conceptual thread that runs through the earlier articles
on wave propagation in SH systems and Ref.~\citep{DKim:PRL22_Bragg}
is that multiple wave scattering is suppressed when long-wavelength
fluctuations are absent from the disorder configuration. In a classical
disordered medium such as an inhomogeneous dielectric material with
a position-dependent permittivity $\epsilon(\bm{r})$,~\citep{RFrank:PRA11_Scalar,ASheremet:PRA20_Absorption}
the control of scattering is realized by modulating the Fourier components
of $\epsilon(\bm{r})$ such that its reciprocal-space spectrum conforms
to a particular distribution. Because multiple scattering underlies
the wave transport phenomena of diffusion and Anderson localization,~\citep{PSheng:Book06_IntroductionWave}
this control of scattering can potentially allow us to engineer the
spectrum of waves  transmitted through the disordered medium. We remark
here that there is a slight difference between the complementary approaches
of Ref.~\citep{DKim:PRL22_Bragg} and existing work on SH systems.~\citep{STorquato:PhysRep18_Hyperuniform,STorquato:PRE16_Hyperuniformity}
The former proposes the reciprocal-space loci of the Fourier components
of the disorder while the latter determines where the Fourier components
should be excluded. 

In our paper, we propose a systematic approach  to engineer disorder
for the selective suppression of scattering to allow the robust transmission
of waves through the disordered medium for isotropic and selective
transparency with orientation-dependent frequency-space windows. Although
it is known that SH systems can act as isotropic low-pass filters,~\citep{OLeseur:Optica16_High,STorquato:PhysRep18_Hyperuniform,STorquato:PRX21_Nonlocal}
we go beyond the current state of the art through the introduction
of a more elaborate substructure in the disorder spectrum, perhaps
foreshadowed by the notion of directional hyperuniformity in Ref.~\citep{STorquato:PRE16_Hyperuniformity},
and show, in the context of a scalar wave model, how disorder can
be more precisely engineered to create mid-band gaps and orientation-dependent
frequency-space transmission windows. We discuss its theoretical basis
by analyzing how the Fourier-space distribution of the disorder  affects
wave scattering  within the framework of the perturbative expansion
of the disordered Green's function and the incident wave function.
A connection is also made with wave transparency in SH systems. Our
analysis sheds light on the relationship between the disorder components
and the on-shell scattering contributions to the disordered Green's
function, and predicts which plane waves are suppressed by disorder
scattering. We use the insights from the analysis to discuss the conditions
for isotropic and selective transparency. For validation, we compute
the coherent transmission coefficient $t(\bm{k})$~\citep{PWAnderson:PMB85_Question}
for a wide range of incident angles and frequencies, using the Atomistic
Green's Function (AGF) method adapted from Refs.~\citep{ZYOng:JAP18_Tutorial,ZYOng:PRB18_Atomistic},
and obtain excellent agreement between the theory and simulation results.
We demonstrate with an 2D example of how the disorder  components
can be combined to suppress wave scattering and to enable robust coherent
transmission for certain plane-wave states at specific incident angles
and frequencies. Possible experimental realizations are also suggested.

\section{Theory of disorder scattering of scalar waves}

\subsection{Scalar wave model}

To discuss the scattering of a harmonic wave $\psi(\bm{r})$ of angular
frequency $\omega$ at position $\bm{r}$ in a disordered medium,
we use the Helmholtz wave equation 
\begin{equation}
\left[\nabla^{2}+\left(\frac{\omega}{c}\right)^{2}\epsilon(\bm{r})\right]\psi(\bm{r})=0\label{eq:WaveEquation}
\end{equation}
where $\nabla^{2}$ and $c$ denote the 2D Laplace operator and wave
speed, respectively. Equation~(\ref{eq:WaveEquation}) has been used
to model transverse magnetic (TM) wave scattering in dielectric materials.~\citep{MDavy:NatCommun15_Universal,LFroufePerez:PNAS17_Band,ABrandstotter:PNAS19_Shaping}
For the purpose of this paper, we interpret the wave function $\psi(\bm{r})$
as the out-of-plane electric field component in a 2D dielectric medium~\citep{MDavy:NatCommun15_Universal,LFroufePerez:PNAS17_Band,ABrandstotter:PNAS19_Shaping}
although our results can be generalized to non-photonic systems. The
static disorder is described by the permittivity function 
\[
\epsilon(\bm{r})=\epsilon_{0}[1+f(\bm{r})]\ ,
\]
where $\epsilon_{0}=1$ denotes the permittivity of the disorder-free
medium; $f(\bm{r})$, which denotes the \emph{disorder function} corresponding
to the position-dependent fluctuations of the permittivity, can be
written as a sum of $N$ Fourier components like in Eq.~(\ref{eq:BerryPotential}),
i.e., 
\begin{equation}
f(\bm{r})=\alpha\sqrt{\frac{2}{N}}\sum_{j=1}^{N}\cos(\bm{p}_{j}\cdot\bm{r}+\phi_{j})\ ,\label{eq:DisorderFunction}
\end{equation}
with the normalization constraint $\lim_{\Omega\rightarrow\infty}\frac{1}{\Omega}\int_{\Omega}d\bm{r}|f(\bm{r})|^{2}=\alpha^{2}$
where $\Omega$ is the area of integration. The dimensionless constant
$\alpha$ is the root-mean-square value of $f(\bm{r})$ and determines
the relative disorder strength in Eq.~(\ref{eq:WaveEquation}) as
well as the `coupling constant' in the Dyson expansion which in our
calculations, we set as $\alpha=0.1$. Instead of a discrete sum,
Eq.~(\ref{eq:DisorderFunction}) can also be expressed as an integral
$f(\bm{r})=\alpha\int d\bm{q}\rho(\bm{q})\exp[i(\bm{q}\cdot\bm{r}+\phi_{\bm{q}})]$
where $\rho(\bm{q})$ is the `density of states'. The Fourier transform
of $f(\bm{r})$ is given by $\mathcal{F}(\bm{q})=\frac{1}{h^{2}}\int_{\Omega}d\bm{r}e^{-i\bm{q}\cdot\bm{r}}f(\bm{r})=\alpha(\frac{2\pi}{h})^{2}\sum_{j=1}^{N}\sqrt{\frac{2}{N}}[e^{i\phi_{j}}\delta(\bm{q}-\bm{p}_{j})+e^{-i\phi_{j}}\delta(\bm{q}+\bm{p}_{j})]$
where $h$ is the 2D unit cell spacing. To facilitate our discussion
of scattering, we rewrite Eq.~(\ref{eq:WaveEquation}) as~\citep{RFrank:PRA11_Scalar,ASheremet:PRA20_Absorption}
\begin{equation}
\left[\nabla^{2}+\kappa_{0}(\omega)^{2}+V(\omega,\bm{r})\right]\psi(\bm{r})=0\ ,\label{eq:PerturbationWaveEquation}
\end{equation}
where $\kappa_{0}(\omega)=\omega\sqrt{\epsilon_{0}}/c$ is the frequency-dependent
wave number for a plane-wave state in the disorder-free medium and
$V(\omega,\bm{r})=\kappa_{0}(\omega)^{2}f(\bm{r})$ denotes the perturbation
term which scales linearly with $\alpha$. In the absence of disorder
where $V(\omega,\bm{r})=0$, Eq.~(\ref{eq:PerturbationWaveEquation})
admits the plane-wave solution $\phi_{0}(\bm{r})=\frac{1}{\sqrt{\Omega}}e^{i\bm{k}\cdot\bm{r}}$
which describes a state with the wave vector $\bm{k}$, such that
$|\bm{k}|=\kappa_{0}$.

\subsection{Dyson expansion of the disordered Green's function}

To elucidate the relationship between wave scattering and disorder,
we analyze the retarded Green's function $G^{+}(\omega,\bm{r},\bm{r}^{\prime})$,
which describes the $\omega$-dependent response at point $\bm{r}$
from a source at $\bm{r}^{\prime}$ in the disordered medium~\citep{EEconomou:Book83_Greens,PSheng:Book06_IntroductionWave}
and is used in perturbative expansion of the wave function $\psi(\bm{r})$.
 Although some of the material in the following discussion on the
Dyson expansion can be found in textbooks, we still discuss it in
order to clarify the role of the Fourier components of $V(\omega,\bm{r})$
in scattering and wave transparency. Before we proceed, we give a
qualitative bird's eye view of our approach to determining wave transparency.
In our analysis, we discuss how $V(\omega,\bm{r})$ affects the perturbative
corrections to the Green's function and the incoming plane-wave state,
which are related through the Lippmann-Schwinger equation, and identify
the condition for wave transparency under which these corrections
vanish. For simplicity, we describe these corrections in terms of
their Fourier transforms $\mathcal{V}(\bm{q})$ with respect to the
reciprocal-space coordinates $\bm{k}$ and $\bm{q}$. This sheds light
on the role of $\mathcal{V}(\bm{q})$ in the integrals for the perturbative
corrections. We exploit the chain structure of these integrals to
identify the $\mathcal{V}(\bm{q})$ components that suppress the perturbative
corrections when the former is set to zero. In addition, we show that
when we set $\mathcal{V}(\bm{q})=0$ for these components, the higher-order
perturbative corrections beyond the Born approximation are also suppressed. 

We remark that in our treatment of the disordered Green's function
$G^{+}$, we do not rely on any kind of configuration averaging and
instead rely on the analysis of wave scattering for each disorder
configuration corresponding to a unique set of phases $\{\phi_{j}\}$
from Eq.~(\ref{eq:DisorderFunction}). This greatly simplifies our
analysis because it eliminates the need to evaluate the configuration-averaged
products of $V$ in the higher-order perturbative corrections to $G^{+}$.~\citep{PSheng:Book06_IntroductionWave}
Without configuration averaging, the integrals for each order in the
perturbative expansion have a chain structure that can be exploited.
This chain structure allows us to show that if that disorder configuration
satisfies the condition $\mathcal{V}(\bm{q})=0$ for an identified
set of $\bm{q}$ vectors such that the lowest-order correction to
$G^{+}$ vanishes, then all the higher-order corrections also vanish
and we can determine the condition for wave transparency. 

To quantify the $O(\alpha^{n})$ perturbative scattering correction
to the incident $\bm{k}$ plane-wave state, we introduce the function
$\mathcal{P}_{n}(\bm{k},\bm{r})$, which is also used in the perturbative
expansion of $G^{+}$, i.e., 
\begin{equation}
\psi(\bm{r})=\phi_{0}(\bm{r})[1+\mathcal{P}_{1}(\bm{k},\bm{r})+\mathcal{P}_{2}(\bm{k},\bm{r})+\ldots]\ ,\label{eq:WaveFunctionPCorrectionSeries}
\end{equation}
and analyze its relationship to $V(\omega,\bm{r})$. The connection
between $\psi(\bm{r})$ and $\phi_{0}(\bm{r})$ can be realized through
the Lippmann-Schwinger equation,~\citep{EEconomou:Book83_Greens}
\[
\psi(\bm{r})=\phi_{0}(\bm{r})+\int d\bm{r}^{\prime}G_{0}^{+}(\omega,\bm{r},\bm{r}^{\prime})V(\omega,\bm{r}^{\prime})\psi(\bm{r}^{\prime})
\]
which we can rewrite as 
\begin{equation}
\psi(\bm{r})=\phi_{0}(\bm{r})+\int d\bm{r}^{\prime}G^{+}(\omega,\bm{r},\bm{r}^{\prime})V(\omega,\bm{r}^{\prime})\phi_{0}(\bm{r}^{\prime})\label{eq:LippmannSchwingerEquation}
\end{equation}
where the second term on the right represents the correction to the
unperturbed state after the perturbation $V$ is switched on. The
significance of $\mathcal{P}_{n}(\bm{k},\bm{r})$ is that it depends
on the scattering strength of the plane wave by the disorder $V$.
Hence, the relation $\mathcal{P}_{n}(\bm{k},\bm{r})=0$ is important
for determining the condition for wave transparency. 

In the absence of disorder ($\alpha=0$), we define the retarded Green's
function $G_{0}^{+}$~\citep{EEconomou:Book83_Greens,PSheng:Book06_IntroductionWave}
using the equation $\left[\nabla^{2}+\kappa_{0}(\omega)^{2}\right]G_{0}^{+}(\omega,\bm{r},\bm{r}^{\prime})=\delta(\bm{r}-\bm{r}^{\prime})$,
where $\delta(\bm{r}-\bm{r}^{\prime})$ denotes the Dirac delta function.~\footnote{In two dimensions, $G_{0}^{+}(\omega,\bm{r},\bm{r}^{\prime})=-\frac{i}{4}H_{0}^{(1)}(\kappa_{0}|\bm{r}-\bm{r}^{\prime}|)$
where $H_{0}^{(1)}$ denotes the zeroth-order Hankel function of the
first kind.~\citep{EEconomou:Book83_Greens,PSheng:Book06_IntroductionWave}} When disorder is present, the retarded Green's function $G^{+}(\omega,\bm{r},\bm{r}^{\prime})$
is defined by the equation $\left[\nabla^{2}+\kappa_{0}(\omega)^{2}+V(\omega,\bm{r})\right]G^{+}(\omega,\bm{r},\bm{r}^{\prime})=\delta(\bm{r}-\bm{r}^{\prime})$. Unlike
$G_{0}^{+}$,  $G^{+}$ has no closed form but is formally related
to $G_{0}^{+}$ through the Dyson equation $G^{+}(\omega,\bm{r},\bm{r}^{\prime})=G_{0}^{+}(\omega,\bm{r},\bm{r}^{\prime})+\int d\bm{r}_{1}G_{0}^{+}(\omega,\bm{r},\bm{r}_{1})V(\omega,\bm{r}_{1})G^{+}(\omega,\bm{r}_{1},\bm{r}^{\prime})$,~\citep{EEconomou:Book83_Greens,PSheng:Book06_IntroductionWave}
which we can expand as a power series 
\begin{align}
G^{+}(\omega,\bm{r},\bm{r}^{\prime}) & =\sum_{n=0}^{\infty}G_{n}^{+}(\omega,\bm{r},\bm{r}^{\prime})\label{eq:DysonExpansion}
\end{align}
where $G_{n}^{+}$ corresponds to the $O(\alpha^{n})$ correction
to $G^{+}$ and can be expressed as a convolution of $G_{0}^{+}$
and $V$. 

In the following discussion, we drop $\omega$ from the arguments
of $G_{n}^{+}$ and $V$ for the sake of brevity. Assuming that the
disorder is sufficiently weak for the series expansion to be valid,
we have for $n=1$ 
\begin{equation}
G_{1}^{+}(\bm{r},\bm{r}^{\prime})=\int d\bm{r}_{1}G_{0}^{+}(\bm{r},\bm{r}_{1})V(\bm{r}_{1})G_{0}^{+}(\bm{r}_{1},\bm{r}^{\prime})\ ,\label{eq:1stOrderDysonTerm}
\end{equation}
and for $n>1$, in general,
\begin{align}
G_{n}^{+}(\bm{r},\bm{r}^{\prime}) & =\int d\bm{r}_{1}\ldots\int d\bm{r}_{n}G_{0}^{+}(\bm{r},\bm{r}_{1})V(\bm{r}_{1})\nonumber \\
 & \times G_{0}^{+}(\bm{r}_{1},\bm{r}_{2})\ldots V(\bm{r}_{n})G_{0}^{+}(\bm{r}_{n},\bm{r}^{\prime})\ .\label{eq:HigherOrderDysonTerm}
\end{align}
We note that for $n\geq1$, the integral in Eq.~(\ref{eq:HigherOrderDysonTerm})
has the chain arrangement  
\begin{equation}
G_{n}^{+}(\bm{r},\bm{r}^{\prime})=\int d\bm{r}_{1}G_{0}^{+}(\bm{r},\bm{r}_{1})V(\bm{r}_{1})G_{n-1}^{+}(\bm{r}_{1},\bm{r}^{\prime})\ .\label{eq:GreensFunctionHierarchy}
\end{equation}
The expression for $G_{n}^{+}$ depends on $G_{n-1}^{+}$ which in
turns depends on $G_{n-2}^{+}$ and so on. Therefore, if we can prove
that $G_{1}^{+}(\bm{r},\bm{r}^{\prime})$, the lowest-order correction
to $G^{+}(\bm{r},\bm{r}^{\prime})$, vanishes for any $\bm{r}$ and
$\bm{r}^{\prime}$ when $V(\bm{r})$ has the right disorder configuration,
then all the higher-order $G_{n}^{+}(\bm{r},\bm{r}^{\prime})$ terms
in Eq.~(\ref{eq:DysonExpansion}) also vanish by induction. For the
isotropic transparency condition, this reduces the problem to a matter
of determining the $V(\bm{r})$ configuration for which $G_{1}^{+}(\bm{r},\bm{r}^{\prime})=0$. 

To simplify the convolution in Eq.~(\ref{eq:1stOrderDysonTerm}),
we use the Fourier transforms $\mathcal{G}_{0}^{+}(\bm{k})$ and $\mathcal{V}(\bm{q})$,
defined by the equations $G_{0}^{+}(\bm{r},\bm{r}^{\prime})=h^{2}\int_{\text{BZ}}\frac{d\bm{k}}{(2\pi)^{2}}e^{i\bm{k}\cdot(\bm{r}-\bm{r}^{\prime})}\mathcal{G}_{0}^{+}(\bm{k})$
and $V(\bm{r})=h^{2}\int_{\text{BZ}}\frac{d\bm{q}}{(2\pi)^{2}}e^{i\bm{q}\cdot\bm{r}}\mathcal{V}(\bm{q})$,
respectively, with  the explicit form of $\mathcal{V}(\bm{q})$  as
$\mathcal{V}(\bm{q})=\kappa_{0}^{2}\mathcal{F}(\bm{q})$. It is well-known
that $\mathcal{V}(\bm{q})$ is proportional to the scattering amplitude
in the Born approximation.~\citep{EEconomou:Book83_Greens} From
Eqs.~(\ref{eq:WaveFunctionPCorrectionSeries}) and (\ref{eq:LippmannSchwingerEquation}),
we obtain the expression 
\begin{equation}
\mathcal{P}_{1}(\bm{k},\bm{r})=h^{4}\int\frac{d\bm{q}}{(2\pi)^{2}}e^{i\bm{q}\cdot\bm{r}}\mathcal{V}(\bm{q})\mathcal{G}_{0}^{+}(\bm{k}+\bm{q})\label{eq:1stPCorrectionTerm}
\end{equation}
which implies that we can rewrite $G_{1}^{+}$ from Eq.~(\ref{eq:1stOrderDysonTerm})
as 
\begin{equation}
G_{1}^{+}(\bm{r},\bm{r}^{\prime})=h^{2}\int\frac{d\bm{k}}{(2\pi)^{2}}e^{i\bm{k}\cdot(\bm{r}-\bm{r}^{\prime})}\mathcal{G}_{0}^{+}(\bm{k})\mathcal{P}_{1}(\bm{k},\bm{r})\ ,\label{eq:1stOrderDysonTermConvolutionIntegral}
\end{equation}
where $\mathcal{G}_{0}^{+}(\bm{k})=\lim_{\eta\rightarrow0^{+}}\frac{1}{h^{2}}\frac{1}{(\kappa_{0}+i\eta)^{2}-|\bm{k}|^{2}}$.
The pole in $\mathcal{G}_{0}^{+}(\bm{k})$ at $|\bm{k}|=\kappa_{0}$~\citep{PSheng:Book06_IntroductionWave}
allows us to simplify Eq.~(\ref{eq:1stOrderDysonTermConvolutionIntegral})
as an angular integral, i.e., 
\begin{equation}
G_{1}^{+}(\bm{r},\bm{r}^{\prime})=\frac{i}{8\pi}\int_{0}^{2\pi}d\theta\exp(i\kappa_{0}\cos\theta|\bm{r}-\bm{r}^{\prime}|)\mathcal{P}_{1}(\kappa_{0}\bm{\hat{k}},\bm{r})\ ,\label{eq:GreensFunctionAngularIntegral}
\end{equation}
where $\theta$ is the angle of $\bm{k}$ with respect to $\bm{r}-\bm{r}^{\prime}$
and $\bm{\hat{k}}$ denotes the unit vector parallel to $\bm{k}$,
and Eq.~(\ref{eq:GreensFunctionAngularIntegral}) implies that $G_{1}^{+}$
depends on the value of $\mathcal{P}_{1}(\bm{k},\bm{r})$ over the
kinematically constrained $|\bm{k}|=\kappa_{0}$ `shell'. Similarly,
we can express Eq.~(\ref{eq:HigherOrderDysonTerm}) as 
\begin{equation}
G_{n}^{+}(\bm{r},\bm{r}^{\prime})=h^{2}\int\frac{d\bm{k}}{(2\pi)^{2}}e^{i\bm{k}\cdot(\bm{r}-\bm{r}^{\prime})}\mathcal{G}_{0}^{+}(\bm{k})\mathcal{P}_{n}(\bm{k},\bm{r})\ ,\label{eq:HigherOrderDysonTermConvolutionIntegral}
\end{equation}
where 
\begin{align}
\mathcal{P}_{n}(\bm{k},\bm{r}) & =h^{4n}\int\frac{d\bm{q}_{1}}{(2\pi)^{2}}\ldots\int\frac{d\bm{q}_{n}}{(2\pi)^{2}}e^{i(\bm{q}_{1}+\ldots+\bm{q}_{n})\cdot\bm{r}}\mathcal{V}(\bm{q}_{1})\nonumber \\
 & \times\mathcal{G}_{0}^{+}(\bm{k}+\bm{q}_{1})\ldots\mathcal{V}(\bm{q}_{n})\mathcal{G}_{0}^{+}(\bm{k}+\bm{q}_{1}+\ldots+\bm{q}_{n})\ ,\label{eq:ExpandedPCorrectionTerm}
\end{align}
which we can rewrite as 
\begin{align}
\mathcal{P}_{n}(\bm{k},\bm{r}) & =h^{4}\int\frac{d\bm{q}_{1}}{(2\pi)^{2}}e^{i\bm{q}_{1}\cdot\bm{r}}\mathcal{V}(\bm{q}_{1})\mathcal{G}_{0}^{+}(\bm{k}+\bm{q}_{1})\nonumber \\
 & \times\mathcal{P}_{n-1}(\bm{k}+\bm{q}_{1},\bm{r})\ ,\label{eq:BranchPCorrectionTerm}
\end{align}
with $\mathcal{P}_{0}=1$. The identity in Eq.~(\ref{eq:BranchPCorrectionTerm})
relates $\mathcal{P}_{n}$ to $\mathcal{P}_{n-1}$ for $n\geq1$ and
like in Eq.~(\ref{eq:GreensFunctionHierarchy}), has a chain arrangement
within the integrand in which $\mathcal{P}_{n}(\bm{k},\bm{r})$ depends
on $\mathcal{P}_{n-1}(\bm{k}+\bm{q}_{1},\bm{r})$ over the same shell
where $|\bm{k}|=|\bm{k}+\bm{q}_{1}|=\kappa_{0}$. This will be useful
for understanding why all higher-order corrections to the $\bm{k}$
plane-wave state are suppressed for $|\bm{k}|<q_{\text{c}}/2$ when
the disorder is described by Eq.~(\ref{eq:DisorderFunction}). 

\subsection{On-shell scattering in perturbative expansion}

The expression in Eq.~(\ref{eq:ExpandedPCorrectionTerm}) lends itself
to an intuitive physical interpretation when it is fully expanded
in $\mathcal{G}_{0}^{+}$ and $\mathcal{V}$. If we confine ourselves
to the first-order (Born) approximation or $\psi(\bm{r})\approx\phi_{0}(\bm{r})[1+\mathcal{P}_{1}(\bm{k},\bm{r})]$,
we can interpret $\mathcal{P}_{1}(\bm{k},\bm{r})$ from Eq.~(\ref{eq:1stPCorrectionTerm})
as the perturbative correction in $\psi(\bm{r})$ due to all the possible
$\bm{k}\stackrel{\mathcal{V}(\bm{q}_{1})}{\longrightarrow}\bm{k}+\bm{q}_{1}$
scattering processes in which $\mathcal{V}(\bm{q}_{1})$ determines
the strength of the scattering process and $\bm{k}$ and $\bm{k}+\bm{q}_{1}$
are the wave vectors of the incident and outgoing plane-wave states,
respectively. Similarly, we can interpret $\mathcal{P}_{2}(\bm{k},\bm{r})=h^{8}\int\frac{d\bm{q}_{1}}{(2\pi)^{2}}\int\frac{d\bm{q}_{2}}{(2\pi)^{2}}e^{i(\bm{q}_{1}+\bm{q}_{2})\cdot\bm{r}}\mathcal{V}(\bm{q}_{1})\mathcal{G}_{0}^{+}(\bm{k}+\bm{q}_{1})\mathcal{V}(\bm{q}_{2})\mathcal{G}_{0}^{+}(\bm{k}+\bm{q}_{1}+\bm{q}_{2})$
as the next-order perturbative correction in $\psi(\bm{r})$ due to
all the possible $\bm{k}\stackrel{\mathcal{V}(\bm{q}_{1})}{\longrightarrow}\bm{k}+\bm{q}_{1}\stackrel{\mathcal{V}(\bm{q}_{2})}{\longrightarrow}\bm{k}+\bm{q}_{1}+\bm{q}_{2}$
scattering processes, and likewise for the remaining $\mathcal{P}_{n}(\bm{k},\bm{r})$
terms where $\bm{k}\stackrel{\mathcal{V}(\bm{q}_{1})}{\longrightarrow}\bm{k}+\bm{q}_{1}\stackrel{\mathcal{V}(\bm{q}_{2})}{\longrightarrow}\ldots\stackrel{\mathcal{V}(\bm{q}_{n})}{\longrightarrow}\bm{k}+\bm{q}_{1}+\ldots+\bm{q}_{n}$.
Because $\mathcal{G}_{0}^{+}(\bm{k})$ has poles at $|\bm{k}|=\kappa_{0}$,~\citep{PSheng:Book06_IntroductionWave}
the presence of the $\mathcal{G}_{0}^{+}(\bm{k})$, $\mathcal{G}_{0}^{+}(\bm{k}+\bm{q}_{1})$,
$\ldots$, $\mathcal{G}_{0}^{+}(\bm{k}+\bm{q}_{1}+\ldots+\bm{q}_{n})$
terms in the integrand of Eq.~(\ref{eq:ExpandedPCorrectionTerm})
implies that the contributions to $\mathcal{P}_{n}(\bm{k},\bm{r})$
for $n\geq1$ are maximized when the wave vectors of the virtual states
($\bm{k}$, $\bm{k}+\bm{q}_{1}$, $\ldots$, $\bm{k}+\bm{q}_{1}+\ldots+\bm{q}_{n}$)
are kinematically restricted to the circular frequency shell of radius
$\kappa_{0}$, i.e., 
\begin{equation}
|\bm{k}|=|\bm{k}+\bm{q}_{1}|=\ldots=|\bm{k}+\bm{q}_{1}+\ldots+\bm{q}_{n}|=\kappa_{0}\ .\label{eq:OnShellCondition}
\end{equation}
The singularity of $\mathcal{G}_{0}^{+}$ implies that the integral
associated with $\mathcal{P}_{n}(\bm{k},\bm{r})$ in Eq.~(\ref{eq:ExpandedPCorrectionTerm})
is limited to scattering between these on-shell states. The kinematic
constraint in Eq.~(\ref{eq:OnShellCondition}) also implies that
the arbitrarily large momentum transfers ($\Delta\bm{k}=\bm{q}_{1}+\ldots+\bm{q}_{n}$)
from multiple scattering are not possible. 

\subsection{Condition for wave transparency}

If a plane-wave state with wave vector $\bm{k}$ propagates through
the disordered medium without attenuation, it means that the medium
is transparent and $\psi(\bm{r})=\phi_{0}(\bm{r})$ because of the
absence of scattering, i.e., $\mathcal{P}_{n}(\bm{k},\bm{r})=0$ for
$n\geq1$. To determine the condition for the absence of scattering,
we analyze the structure of the integrand in $\mathcal{P}_{n}(\bm{k},\bm{r})$
from Eq.~(\ref{eq:ExpandedPCorrectionTerm}) and find the configuration
of the disorder $V(\bm{r})$, in the form of the $\mathcal{V}(\bm{q})=0$
spectrum, that is compatible with $\mathcal{P}_{n}(\bm{k},\bm{r})=0$.

\subsubsection{Isotropic transparency}

In the case of \emph{isotropic} transparency like in SH systems, we
have to find the $\mathcal{V}(\bm{q})$ spectrum compatible with $\mathcal{P}_{n}(\bm{k},\bm{r})=0$
for \emph{all} $\bm{k}$'s that satisfy $|\bm{k}|=\kappa_{0}$ and
Eq.~(\ref{eq:HigherOrderDysonTermConvolutionIntegral}) implies the
more stringent condition $G^{+}(\bm{r},\bm{r}^{\prime})\approx G_{0}^{+}(\bm{r},\bm{r}^{\prime})$.
This problem is however greatly simplified given the chain structure
of Eq.~(\ref{eq:GreensFunctionHierarchy}) which suggests that we
need only to determine the $\mathcal{V}(\bm{q})$ spectrum corresponding
to $G_{1}^{+}(\bm{r},\bm{r}^{\prime})=0$ because Eq.~(\ref{eq:GreensFunctionHierarchy})
implies that $G_{n}^{+}(\bm{r},\bm{r}^{\prime})=0$ for $n>1$. 

The structure of the angular integral in Eq.~(\ref{eq:GreensFunctionAngularIntegral})
means that $G_{1}^{+}(\bm{r},\bm{r}^{\prime})$ depends on $\mathcal{P}_{1}(\bm{k},\bm{r})$
distributed over all $\bm{k}$'s that satisfy $|\bm{k}|=\kappa_{0}$.
 In this angular integral, the value of $\mathcal{P}_{1}(\bm{k},\bm{r})$
on the shell depends on the magnitude of $\mathcal{V}(\bm{q})$ for
$|\bm{q}|<2\kappa_{0}$, which we physically interpret as a bottleneck
limiting the availability of the on-shell $\bm{k}\rightarrow\bm{k}+\bm{q}$
scattering phase space for $|\bm{k}|=|\bm{k}+\bm{q}|=\kappa_{0}$
.  Hence, if we wish to make the approximation $G_{1}^{+}(\bm{r},\bm{r}^{\prime})\approx0$,
we should set $\mathcal{V}(\bm{q})=0$ for $|\bm{q}|<2\kappa_{0}$
to suppress this on-shell scattering contribution to $G_{1}^{+}(\bm{r},\bm{r}^{\prime})$.
This is similar to the condition $S(\bm{q})=0$ for $|\bm{q}|<q_{\text{SH}}$
in SH systems which are also transparent for incoming plane-wave states
that satisfy $|\bm{k}|<q_{\text{SH}}/2$.~\citep{OLeseur:Optica16_High,JKim:NJP20_Effective,STorquato:PRX21_Nonlocal}
In our case, we may regard $|\mathcal{V}(\bm{q})|^{2}$ as the analog
of $S(\bm{q})$ for $|\bm{q}|>0$ and like in SH systems, define a
finite `exclusion zone' centered around the origin in reciprocal space
for the nonzero $\mathcal{V}(\bm{q})$ components. 

\begin{figure}
\begin{centering}
\includegraphics[width=8cm]{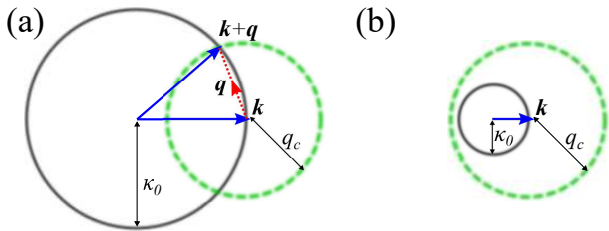}
\par\end{centering}
\caption{The black Ewald circle is the shell containing the loci of all possible
$\bm{k}$ states such that $|\bm{k}|=\kappa_{0}$. The green circle
of radius $q_{\text{c}}$ is centered at $\bm{k}$ and contains the
loci of all the possible $\bm{q}$ disorder Fourier components in
the Berry potential. (a) If $\kappa_{0}>q_{\text{c}}/2$, then the
on-shell $\bm{k}\rightarrow\bm{k}+\bm{q}$ scattering process is allowed
because the two circles intersect. (b) If $\kappa_{0}<q_{\text{c}}/2$,
then the two circles do not intersect and there is no on-shell scattering.}

\label{fig:EwaldCircles}
\end{figure}

 Therefore, if we set $V(\bm{r})$ from Eq.~(\ref{eq:PerturbationWaveEquation})
to be proportional to $U(\bm{r};\{\phi_{j}\})$ from Eq.~(\ref{eq:BerryPotential})
such that the  Fourier components in $f(\bm{r})$ are only non-zero
when $|\bm{q}|=q_{\text{c}}$, then we can approximate $G^{+}(\omega,\bm{r},\bm{r}^{\prime})\approx G_{0}^{+}(\omega,\bm{r},\bm{r}^{\prime})$
for $\omega<\omega_{\text{c}}$ where $\omega_{c}=\frac{cq_{\text{c}}}{2\sqrt{\epsilon_{0}}}$
is the cutoff frequency. This implies that any incident plane wave
with frequency $\omega<\omega_{\text{c}}$, or equivalently with wave
number $|\bm{k}|<q_{\text{c}}/2$, can propagate through the disordered
region with near total transparency, consistent with the principal
finding of Ref.~\citep{DKim:PRL22_Bragg}. Similar conditions for
the cutoff frequency in SH systems have also been derived in Ref.~\citep{OLeseur:Optica16_High}
and also by Torquato and Kim~\citep{JKim:NJP20_Effective,STorquato:PRX21_Nonlocal}.
  Figure~\ref{fig:EwaldCircles} shows the geometrical interpretation
for the on-shell scattering contribution in $\mathcal{P}_{1}(\bm{k},\bm{r})$.
We draw an Ewald circle of all the possible $|\bm{k}|=\kappa_{0}$
states and another circle containing the disorder modes. If $\omega>\omega_{\text{c}}$,
then the Ewald circle is large enough for the two circles to intersect
and disorder scattering is allowed. If $\omega<\omega_{\text{c}}$,
no disorder scattering is allowed.

\subsubsection{Selective transparency}

Beyond isotropic transparency, we can also fine-tune $\mathcal{V}(\bm{q})$
to generate wave transparency for a set of $\bm{k}$ plane-wave states
that is more selective than the $0<\omega<\omega_{\text{c}}$ frequency
band in isotropic transparency. Instead of setting $\mathcal{V}(\bm{q})=0$
in the entire $|\bm{q}|<q_{\text{c}}$ neighborhood, we limit the
$\mathcal{V}(\bm{q})=0$ condition to a subset of the Fourier components
in the $|\bm{q}|<q_{\text{c}}$ region to generate a smaller window
of wave transparency containing the selected $\bm{k}$ plane-wave
states. In other words, we permit some non-zero Fourier components
in the `exclusion zone' to interact (scatter) with the incoming plane-wave
states that are outside of this window. To do this, we need to determine
the condition for $\mathcal{P}_{n}(\bm{k},\bm{r})=0$ for the $\bm{k}$
states in this window. We first show how this is determined for $\mathcal{P}_{1}(\bm{k},\bm{r})$
in Eq.~(\ref{eq:1stPCorrectionTerm}), which is associated with the
$\bm{k}\stackrel{\mathcal{V}(\bm{q}_{1})}{\longrightarrow}\bm{k}+\bm{q}_{1}$
scattering processes of an individual $\bm{k}$ state, as the procedure
can be generalized to higher-order $\mathcal{P}_{n}(\bm{k},\bm{r})$
terms because the integrand for $\mathcal{P}_{n}(\bm{k},\bm{r})$
in Eq.~(\ref{eq:BranchPCorrectionTerm}) also contains the $\mathcal{V}(\bm{q})\mathcal{G}_{0}^{+}(\bm{k}+\bm{q})$
term which determines the loci of $\bm{q}$ for the $\mathcal{V}(\bm{q})$
contributing to the integral. 

Given the presence of the $\mathcal{V}(\bm{q})\mathcal{G}_{0}^{+}(\bm{k}+\bm{q})$
term in the integrand, which we associate with the $\bm{k}\rightarrow\bm{k}+\bm{q}$
scattering process shown in Fig.~\ref{fig:EwaldCircles}(a), we minimize
the integral in Eq.~(\ref{eq:1stPCorrectionTerm}) and hence $\mathcal{P}_{1}(\bm{k},\bm{r})$
by setting $\mathcal{V}(\bm{q})=0$ for all possible values of $\bm{q}$
that satisfy $|\bm{k}+\bm{q}|=\kappa_{0}$. These loci of $\bm{q}$
values for $\mathcal{V}(\bm{q})=0$ comprise a circle of radius $\kappa_{0}$
centered at $\bm{q}=-\bm{k}$ in reciprocal space and contain all
the Fourier components that can interact with that particular $\bm{k}$
state. For each $\bm{k}$ state, we have one circle. We note that
this circle is contained within the $|\bm{q}|<2\kappa_{0}$ neighborhood.
Hence, to generate transparency for a window of $\bm{k}$ states,
we set the $\mathcal{V}(\bm{q})=0$ condition over the reciprocal-space
region defined by the superposition of these circles. If this transparency
window includes all the $\bm{k}$ states with $|\bm{k}|<q_{\text{c}}/2$
like in SH systems, then, as expected, we have to impose the $\mathcal{V}(\bm{q})=0$
condition over the entire $|\bm{q}|<q_{\text{c}}$ neighborhood. Otherwise,
the $\mathcal{V}(\bm{q})=0$ condition has to be applied to only a
subregion of the $|\bm{q}|<q_{\text{c}}$ neighborhood. This procedure
also minimizes the higher-order $\mathcal{P}_{n}(\bm{k},\bm{r})$
terms, which are associated with the $\bm{k}\stackrel{\mathcal{V}(\bm{q}_{1})}{\longrightarrow}\bm{k}+\bm{q}_{1}\stackrel{\mathcal{V}(\bm{q}_{2})}{\longrightarrow}\ldots\stackrel{\mathcal{V}(\bm{q}_{n})}{\longrightarrow}\bm{k}+\bm{q}_{1}+\ldots+\bm{q}_{n}$
process for multiple scattering, because it suppresses the $\bm{k}\stackrel{\mathcal{V}(\bm{q}_{1})}{\longrightarrow}\bm{k}+\bm{q}_{1}$
part of the scattering process.

This result can be reached from another perspective by considering
the self-energy, which describes the frequency shift and inverse lifetime
caused by the disorder scattering, as defined by $\Sigma(\omega,\bm{k})=\mathcal{G}_{0}^{+}(\bm{k})^{-1}-\langle\mathcal{G}(\bm{k})\rangle^{-1}$
where $\langle\mathcal{G}(\bm{k})\rangle$ is the configuration-averaged
disordered Green's function in reciprocal space. To the lowest non-zero
$O(\alpha^{2})$ approximation, we can write~\citep{GDMahan:Book00_Many}
\begin{align}
\Sigma(\omega,\bm{k}) & \approx\int\frac{d\bm{q}}{(2\pi)^{2}}\langle\mathcal{V}(\bm{q})\mathcal{V}(-\bm{q})\rangle\mathcal{G}_{0}^{+}(\bm{k}+\bm{q})\nonumber \\
 & \approx\int\frac{d\bm{q}}{(2\pi)^{2}}|\mathcal{V}(\bm{q})|^{2}\mathcal{G}_{0}^{+}(\bm{k}+\bm{q})\label{eq:DisorderSelfEnergy}
\end{align}
since $\mathcal{V}(-\bm{q})=\mathcal{V}(\bm{q})^{\dagger}$. Given
the singularity in $\mathcal{G}_{0}^{+}(\bm{k}+\bm{q})$ and the linear
dispersion $\omega(\bm{k})=c|\bm{k}|$, the $\bm{q}$ integral for
$\Sigma(\omega,\bm{k})$ is effectively taken over a circular shell
of radius $\kappa_{0}$ centered at $\bm{q}=-\bm{k}$ in reciprocal
space, like in the integral for $\mathcal{P}_{1}(\bm{k},\bm{r})$
in Eq.~(\ref{eq:1stPCorrectionTerm}). Therefore, if we set $\mathcal{V}(\bm{q})=0$
on this shell and use the on-shell approximation $|\bm{k}|=\kappa_{0}$,
then we have $\Sigma(\omega,\bm{k})\approx0$ which we interpret as
the absence of scattering for the $\bm{k}$ plane-wave state.

\section{Coherent transmission simulation and analysis}

\subsection{Simulation setup}

To validate our analysis, we compute the \emph{coherent} transmission
coefficient $t(\bm{k})$,~\citep{PWAnderson:PMB85_Question} which
determines the proportion of the wave flux passing through the disordered
region without blurring, for a range of $\bm{k}$ states. If $|t(\bm{k})|^{2}=1$,
the disordered medium is completely transparent. We approximate Eq.~(\ref{eq:WaveEquation})
using a 2D square lattice~\citep{PSheng:Book06_IntroductionWave},
in which the second derivatives are replaced by finite differences,
i.e., $\nabla^{2}\psi(\bm{r})\approx\frac{1}{a^{2}}[-4\psi(\bm{r})+\psi(\bm{r}+a\hat{\bm{x}})+\psi(\bm{r}-a\hat{\bm{x}})-\psi(\bm{r}+a\hat{\bm{y}})+\psi(\bm{r}-a\hat{\bm{y}})]$
where $a$ is the 2D lattice constant determined by the length scale
of the scattering problem. The resulting eigenvalue equation $\frac{1}{a^{2}}[4\psi(\bm{r})-\psi(\bm{r}+a\hat{\bm{x}})-\psi(\bm{r}-a\hat{\bm{x}})-\psi(\bm{r}+a\hat{\bm{y}})-\psi(\bm{r}-a\hat{\bm{y}})]=\omega^{2}\frac{\epsilon(\bm{r})}{c^{2}}\psi(\bm{r})$
can be written in the matrix form $\bm{K}\Psi=\omega^{2}\bm{M}\Psi$,
where $\bm{K}$ is the finite-difference matrix, $\Psi$ is a column
vector with $\psi(\bm{r})$ as its vector elements, and $\bm{M}$
is a diagonal matrix with $\frac{\epsilon(\bm{r})}{c^{2}}$ as its
diagonal elements. This formulation sets us up for the direct scattering
amplitude calculations using the AGF method,~\citep{ZYOng:JAP18_Tutorial,ZYOng:PRB18_Atomistic}
a technique developed to study phonon scattering.~\citep{ZYOng:PRB20_Structure,ZYOng:EPL21_Specular}
In the disorder-free lattice where $\epsilon(\bm{r})=\epsilon_{0}$,
the dispersion relationship is given by~\citep{PSheng:Book06_IntroductionWave}
$\omega(\bm{k})^{2}=\frac{4c^{2}}{a^{2}}[\sin^{2}(\frac{1}{2}k_{x}a)+\sin^{2}(\frac{1}{2}k_{y}a)]$.
In the continuum ($a\rightarrow0$) limit, we obtain $|\bm{k}|=\kappa_{0}$
and recover the linear $\omega(\bm{k})=c|\bm{k}|$ relationship.~\citep{PSheng:Book06_IntroductionWave} 

\begin{figure}
\begin{centering}
\includegraphics[width=8cm]{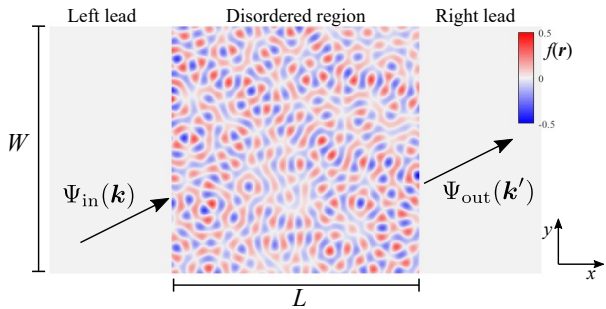}
\par\end{centering}
\caption{Schematic of the simulation setup for the AGF calculations. The disordered
region ($0\protect\leq x\protect\leq L$) is confined between the
two semi-infinite disorder-free leads on the left ($x<0$) and right
($x>L$) where $f(\bm{r})=0$. In the transverse $y$ direction, we
impose periodic boundary conditions. }

\label{fig:SimSetUp}
\end{figure}

Our AGF simulation setup is shown in Fig.~\ref{fig:SimSetUp}. A
disordered region of width $W$ and length $L$ is sandwiched between
the semi-infinite and disorder-free left and right leads where $f(\bm{r})=0$.
We fix $W=200a$ and let $L$ vary. We use the dimensionless variable
$\kappa_{0}a$ to represent the frequency. For each $\bm{k}$ plane
wave, we compute $|t(\bm{k})|^{2}=|S(\bm{k},\bm{k})|^{2}$, where
$S(\bm{k}^{\prime},\bm{k})$ is the scattering amplitude between the
incoming state $\Psi_{\text{in}}(\bm{k})$ and the outgoing state
$\Psi_{\text{out}}(\bm{k}^{\prime})$ at frequency $\omega$. At each
frequency step, we determine all the kinematically allowed $\bm{k}$
states and compute their $S(\bm{k}^{\prime},\bm{k})$ and $|t(\bm{k})|^{2}$
values. For consistency with our analysis based on the continuum limit,
we restrict the range of $\bm{k}$ in our AGF calculations to $|\bm{k}|\apprle1/a$
or $\kappa_{0}a\leq1$. 

\subsection{Simulation results}

\subsubsection{Isotropic wave transparency}

\begin{figure}
\begin{centering}
\includegraphics[width=8cm]{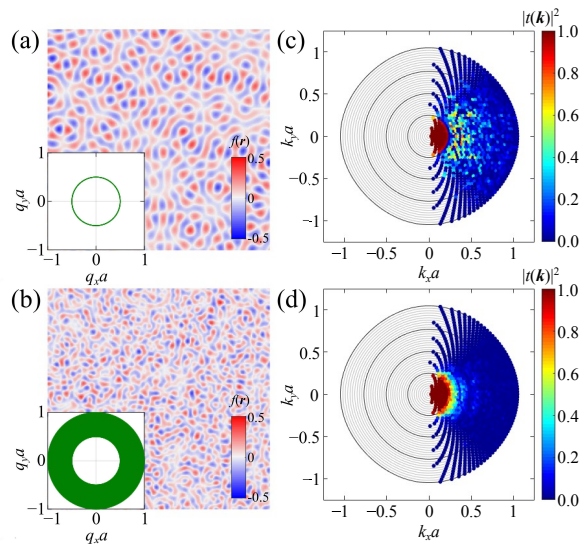}
\par\end{centering}
\caption{The disorder function $f(\bm{r})$ for disorder modes at (a) $|\bm{q}|=0.5/a$
and (b) $0.5/a\protect\leq|\bm{q}|\protect\leq1/a$. The value of
$f(\bm{r})$ is indicated by the color bar in the bottom right corner.
The bottom left corner insets show the loci of the disorder modes
in $\bm{q}$ space. The corresponding coherent transmission $|t(\bm{k})|^{2}$
spectra for $L=5W$ are shown in (c) and (d), respectively, with each
$\bm{k}$ plane wave represented by a colored dot indicating its numerical
value according to the color bar. To guide the eye, we draw the dimensionless
frequency-dependent $\kappa_{0}a$ contour lines at intervals of $0.025$
(gray lines) and $0.25$ (black lines).}

\label{fig:CircularManifoldResults}
\end{figure}

Figure~\ref{fig:CircularManifoldResults}(a) shows an instance of
$f(\bm{r})$ in which its Fourier components are distributed uniformly
over a circular manifold (or `ring') with a radius of $|\bm{q}|=q_{\text{c}}=0.5/a$,
as shown in the inset of Fig.~\ref{fig:CircularManifoldResults}(a).
The corresponding $|t(\bm{k})|^{2}$ spectrum for $L=5W$ is shown
in Fig.~\ref{fig:CircularManifoldResults}(b) for the frequency range
$0<\kappa_{0}a\leq1$. We observe that for $\kappa_{0}>q_{\text{c}}/2$,
$|t(\bm{k})|^{2}\approx0$ for almost every state as a result of multiple
scattering by the disorder. This reduction in $|t(\bm{k})|^{2}$ for
$\kappa_{0}>q_{\text{c}}/2$ is a result of the disordered domain
being much larger than the mean free paths of the plane-wave states
with $\kappa_{0}>q_{\text{c}}/2$. At $\kappa_{0}=q_{\text{c}}/2$,
there is a sharp transition or cutoff frequency, below which $|t(\bm{k})|^{2}\approx1$
or near-perfect transparency for every state regardless of the angle
of incidence. Nonetheless, we observe some speckling in the $|t(\bm{k})|^{2}$
spectrum above the  cutoff frequency due to the random phase in the
disorder modes. If we distribute the disorder  components uniformly
over a band of rings ($0.5/a\leq|\bm{q}|\leq1/a$), as shown in the
inset of Fig.~\ref{fig:CircularManifoldResults}(c), instead of a
single ring, then the speckling is smoothed out, as shown in Fig.~\ref{fig:CircularManifoldResults}(d),
because of the greater range of  $\mathcal{V}(\bm{q})$ components
available for on-shell scattering. 

\subsubsection{Orientation and frequency-dependent transparency window \label{subsec:TransparencyWindow}}

\begin{figure*}
\begin{centering}
\includegraphics[width=12cm]{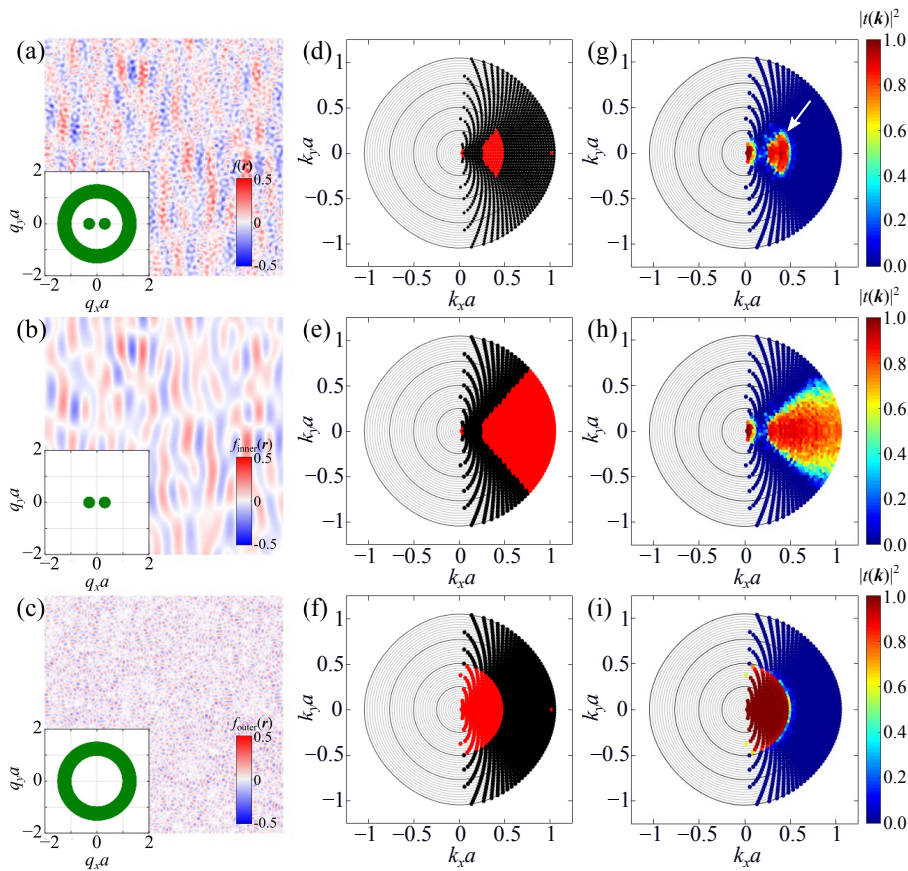}
\par\end{centering}
\caption{Plot of the disorder function for (a) $f(\bm{r})$, (b) $f_{\text{inner}}(\bm{r})$,
and (c) $f_{\text{outer}}(\bm{r})$. The value of $f(\bm{r})$ is
indicated by the color bar in the bottom right corner. The bottom
left corner insets show the loci of the  Fourier components in $\bm{q}$
space. The corresponding distributions of the scattered (black dots)
and unscattered (red dots) $\bm{k}$ plane-wave states are shown in
(d) to (f) while the $|t(\bm{k})|^{2}$ spectra for $L=20W$ are shown
in (g) to (i) with the numerical values represented by the color bar.
The white arrow in (g) points to the window of transparency for $f(\bm{r})$. }
\label{fig:CompositeDisorderTransmissionSpectra}
\end{figure*}

The smoothing of the speckling in Fig.~\ref{fig:CircularManifoldResults}(d)
implies that the the effects of the disorder  Fourier components are
additive, i.e., by selectively including more  Fourier components
in $f(\bm{r})$ we can introduce more scattering pathways to modify
the $|t(\bm{k})|^{2}$ spectrum by filtering out unwanted $\bm{k}$
states. We go beyond the SH systems by introducing an anisotropic
substructure in the distribution of the Fourier components. We elaborate
on this idea with the example in Fig.~.\ref{fig:CompositeDisorderTransmissionSpectra}.
Figure~\ref{fig:CompositeDisorderTransmissionSpectra}(a) shows the
anisotropic $f(\bm{r})$ distribution ($f=f_{\text{inner}}+f_{\text{outer}}$)
obtained from combining two disorder distributions $f_{\text{inner}}(\bm{r})$
and $f_{\text{outer}}(\bm{r})$, shown in Figs.~\ref{fig:CompositeDisorderTransmissionSpectra}(b)
and (c), respectively. Each of the disorder distributions corresponds
to a set of distinctive scattering pathways. The highly anisotropic
$f_{\text{inner}}$ comprises  Fourier components distributed over
two  circular pockets with radius of $0.2/a$ and centered away from
the origin at $q_{x}=\pm0.3/a$ unlike the examples in Fig.~\ref{fig:CircularManifoldResults}.
On the other hand, $f_{\text{outer}}$, which describes SH-like disorder,
is isotropic and comprises  Fourier components that are centered at
the origin and distributed over a band of rings with radii varying
between $1/a$ and $1.5/a$.

To understand how combining disorder distributions affects the scattering
pathways, we plot in Figs.~\ref{fig:CompositeDisorderTransmissionSpectra}(d)
to (f) the $\bm{k}$ plane-wave states that are unaffected by on-shell
scattering for $f(\bm{r})$, $f_{\text{inner}}(\bm{r})$, and $f_{\text{outer}}(\bm{r})$,
respectively. At each given $\kappa_{0}$, we compute all the possible
$\bm{k}\rightarrow\bm{k}+\bm{q}$ on-shell scattering transitions
for each $\bm{k}$ state. If none of the $\bm{q}$'s lie in the region
occupied by the Fourier components, then that $\bm{k}$ state is
considered unaffected by on-shell scattering. The distribution of
unscattered $\bm{k}$ states for $f_{\text{inner}}$ in Fig.~\ref{fig:CompositeDisorderTransmissionSpectra}(e)
shows a distinctive anisotropy. At more oblique angles of incidence,
the $\bm{k}$ states are more likely to be scattered. The low-frequency
plane waves with $\kappa_{0}a<0.05$ are however unaffected by on-shell
scattering because the disorder component closest to the origin is
at $\bm{q}=(\pm0.1/a,0)$. In the $0.05\leq\kappa_{0}a\leq0.25$ frequency
range, there is a transmission gap\footnote{We define the transmission gap as the frequency range in which plane
waves cannot propagate through the disordered medium.} because of the absence of unscattered plane waves. The upper bound
of this gap is determined by the position of the disorder mode furthest
from the origin at $\bm{q}=(\pm0.5/a,0)$ and can be adjusted by changing
the radius and position of the  circular pockets in $f_{\text{inner}}$.
At higher frequencies ($\kappa_{0}a>0.25$), a range of unscattered
$\bm{k}$ states exists for more acute incident angles because there
are no disorder modes available in $f_{\text{inner}}$ to scatter
the plane wave at higher frequencies for small incident angles. For
$f_{\text{outer}}$, the distribution of unscattered $\bm{k}$ states
in Fig.~\ref{fig:CompositeDisorderTransmissionSpectra}(f) show a
sharp cutoff at $\kappa_{0}a<0.5$ because the disorder Fourier components
are located in the $1/a\leq|\bm{q}|\leq1.5/a$ band.

Hence, for $f$, Fig.~\ref{fig:CompositeDisorderTransmissionSpectra}(d)
shows the distribution of unscattered $\bm{k}$ states, which is equal
to the intersection of the unscattered $\bm{k}$ states in Figs.~\ref{fig:CompositeDisorderTransmissionSpectra}(e)
and (f). Figures~\ref{fig:CompositeDisorderTransmissionSpectra}(g)
to (i) show the $|t(\bm{k})|^{2}$ spectra for $L=20W$ where a larger
$L$ is chosen to magnify the effects of multiple scattering. We see
a close correspondence between Figs.~\ref{fig:CompositeDisorderTransmissionSpectra}(d)
and (g), with the $|t(\bm{k})|^{2}$ values highest for the unscattered
$\bm{k}$ states, validating our theory of how disorder  Fourier components
affect wave scattering and transport. The $|t(\bm{k})|^{2}$ spectrum
for $f$ can also be approximated by taking the product of the $|t(\bm{k})|^{2}$'s
for $f_{\text{inner}}$ and $f_{\text{outer}}$ from Figs.~\ref{fig:CompositeDisorderTransmissionSpectra}(h)
and (i). Figure~\ref{fig:CompositeDisorderTransmissionSpectra}(g)
shows that by combining the  Fourier components of $f_{\text{inner}}$
and $f_{\text{outer}}$ to filter out the unwanted $\bm{k}$ states,
we are able to create transmission gaps and  a window of high transparency
{[}Fig.~\ref{fig:CompositeDisorderTransmissionSpectra}(g){]} in
the $|t(\bm{k})|^{2}$ spectrum, in which incoming plane waves can
be robustly transmitted through the disordered medium without blurring.
The correspondence between Figs.~\ref{fig:CompositeDisorderTransmissionSpectra}(d)
and (g) is however not perfect because of wave reflection at the boundary
between the leads and the disordered region.  Nevertheless, the example
from Fig.~\ref{fig:CompositeDisorderTransmissionSpectra} shows that
we can combine groups of disorder Fourier components to engineer specific
wave transport properties such as the transparency  window; we use
$f_{\text{in}}$ to create anisotropy and a transmission gap and $f_{\text{out}}$to
create a low-pass filter. Furthermore, the orientation, size and position
of the transparency  window can be modified by changing the loci of
the  circular pockets and ring in $f$. In Figs.~\ref{fig:45DegreeDisorderData}
and \ref{fig:90DegreeDisorderData}, we also show how the $|t(\bm{k})|^{2}$
spectrum changes when we modify $f_{\text{inner}}$ by changing the
orientation of the circular pockets. 

\begin{figure}
\includegraphics[scale=0.55]{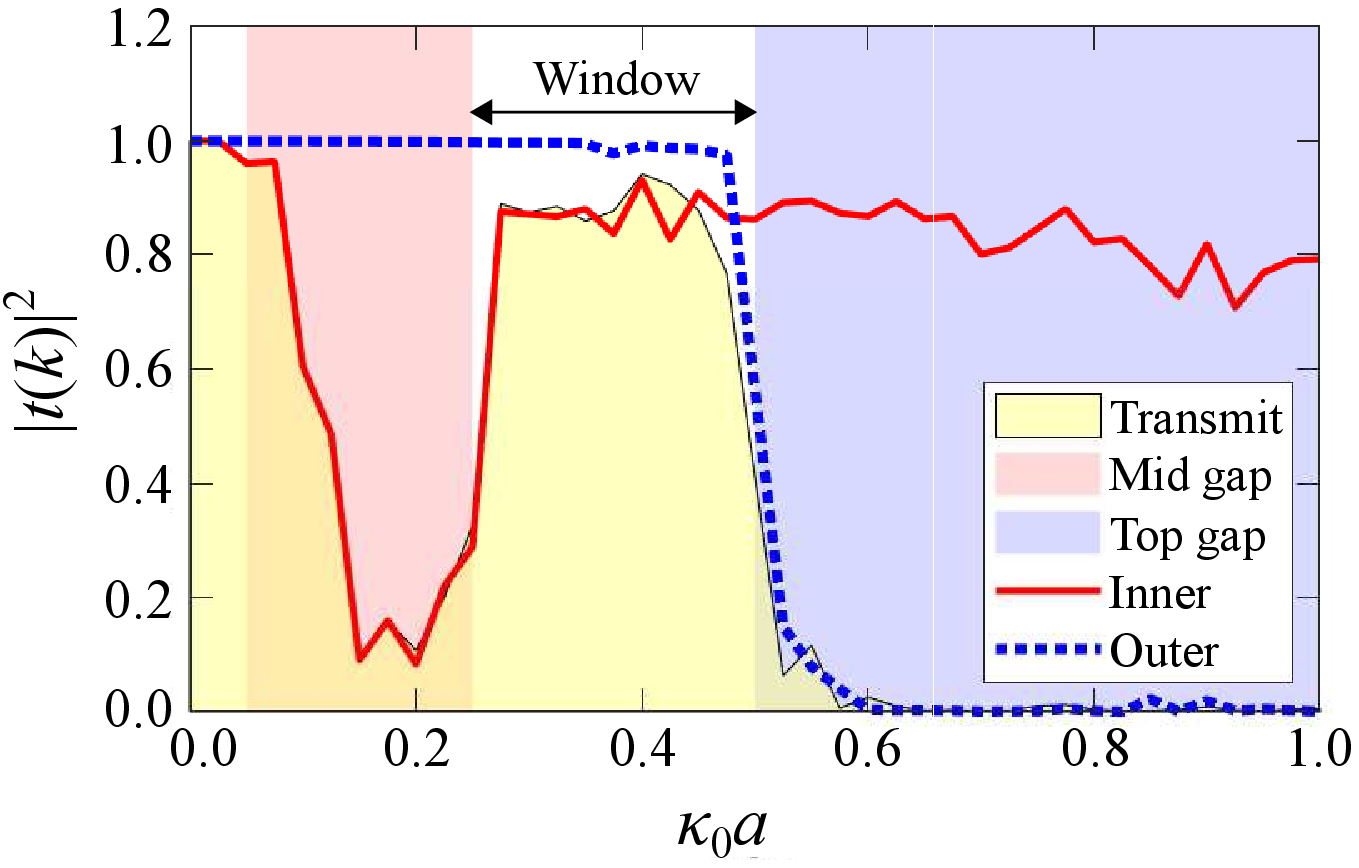}

\caption{Coherent transmission $|t(\bm{k})|^{2}$ spectrum (yellow shaded area
under the curve) for the disorder function $f(\bm{r})$ as a function
of frequency $\kappa_{0}a$ at normal incidence ($k_{y}=0$) with
the label `Transmit'. The spectrum is derived from Fig.~\ref{fig:CompositeDisorderTransmissionSpectra}(g).
The $|t(\bm{k})|^{2}$ spectra for $f_{\text{inner}}$ and $f_{\text{outer}}$
are labeled `Inner' (red solid line) and `Outer' (dashed blue line),
respectively, with the corresponding transmission gaps labeled `Mid
gap' (red shaded rectangle) and `Top gap' (blue shaded rectangle).
The transmission window in the $0.25<\kappa_{0}a<0.5$ range is indicated
by the double-headed arrow (`Window').}

\label{fig:NormalIncidenceTransmissionSpectrum}
\end{figure}

For greater clarity, we plot in Fig.~\ref{fig:NormalIncidenceTransmissionSpectrum}
the coherent transmission $|t(\bm{k})|^{2}$ spectrum from Fig.~\ref{fig:CompositeDisorderTransmissionSpectra}(g)
for incoming plane-wave states at normal incidence ($k_{y}=0$). We
observe two transmission bands \textendash{} one at low frequencies
($\kappa_{0}a<0.05$) and the other in the `window' ($0.25<\kappa_{0}a<0.5$).
The two transmission bands are separated by a `mid gap' ($0.05\leq\kappa_{0}a\leq0.25$)
that originates from scattering by the Fourier components associated
with $f_{\text{inner}}$ while the `window' is bounded from above
by a `top gap' ($\kappa_{0}a\geq0.5$) originating from scattering
by the Fourier components associated with $f_{\text{outer}}$. 

\subsection{Possible experimental realization}

This phenomenon can be simulated for TM waves in a 2D photonic crystal
(PC) with tunable site disorder. A prototype would be a 2D square
lattice of cylindrical dielectric rods~\citep{EIstrate:RMP06_Photonic,KBusch:PR07_Periodic}
with a radius of $R$ and nearest-neighbor distance of $h$. To realize
the \emph{spatial structure} of the disorder described by $f(\bm{r})$,
we can let the cylinder radius be site dependent, i.e., $R(\bm{r})=R_{0}[1+f(\bm{r})]$,
where $R_{0}$ is the average radius of the rods in the photonic crystal,
so that the perturbation is proportional to $f(\bm{r})$. The coherent
transmission spectrum through a disordered PC sandwiched between two
 disorder-free PC leads as in Fig.~\ref{fig:SimSetUp} should be
similar to those in Fig.~\ref{fig:CircularManifoldResults}. Alternatively,
the coherent transmission spectrum should be observed in a disordered
PC with a position-dependent refractive index.~\citep{TSchwartz:Nature07_Transport}

Another possible approach for experimental realization would be to
allow the amplitude of the nonzero Fourier components of $f(\bm{r})$
to vary while we fix the loci of the zero-amplitude Fourier components
in $\bm{q}$ space. For instance, in the example in Section \ref{subsec:TransparencyWindow},
the zero-amplitude Fourier components would be confined to the region
surrounding the two circular pockets and bounded by the ring as shown
in the inset of Fig.~\ref{fig:CompositeDisorderTransmissionSpectra}(a).
This provides us with more flexibility in the design of disordered
media with selective transparency and such disorder configurations
can be realized through collective-coordinate optimization.~\citep{OUche:PRE04_Constraints,STorquato:PhysRep18_Hyperuniform}

\section{Summary}

We have elucidated the role of disorder, in terms of its Fourier components,
 in wave scattering in a 2D disordered medium. We show how the disorder
configuration, as determined by $\mathcal{V}(\bm{q})$, can be engineered
for isotropic and selective wave transparency. Using numerical simulations,
we demonstrate an approach where, by choosing the appropriate combination
of  Fourier components, wave scattering can be selectively suppressed
and the transport properties of the material can be engineered to
create transmission gaps and allow specific incident waves to be robustly
and coherently transmitted in orientation-dependent frequency windows.
This Fourier components-based approach, which requires neither nontrivial
topological wave properties nor a non-Hermitian medium, can be generalized
to tailor the transport properties of disordered media for fundamental
investigations of disordered systems and applications in acoustics,
imaging, photonics, and structural health monitoring. 

\begin{acknowledgments}
I thank Donghwan Kim and Eric Heller of Harvard University for clarifying
some concepts in their paper.~\citep{DKim:PRL22_Bragg} I thank Salvatore
Torquato of Princeton University for drawing my attention to the phenomenon
of wave transparency in stealthy hyperuniform systems.  I acknowledge
support for this work by A{*}STAR, Singapore with funding from the
Polymer Matrix Composites Program (SERC Grant No. A19C9a004).
\end{acknowledgments}

\appendix

\section{Coherent transmission spectra for other combinations of $f_{\text{inner}}$
and $f_{\text{outer}}$ }

\begin{figure*}
\includegraphics[width=12cm]{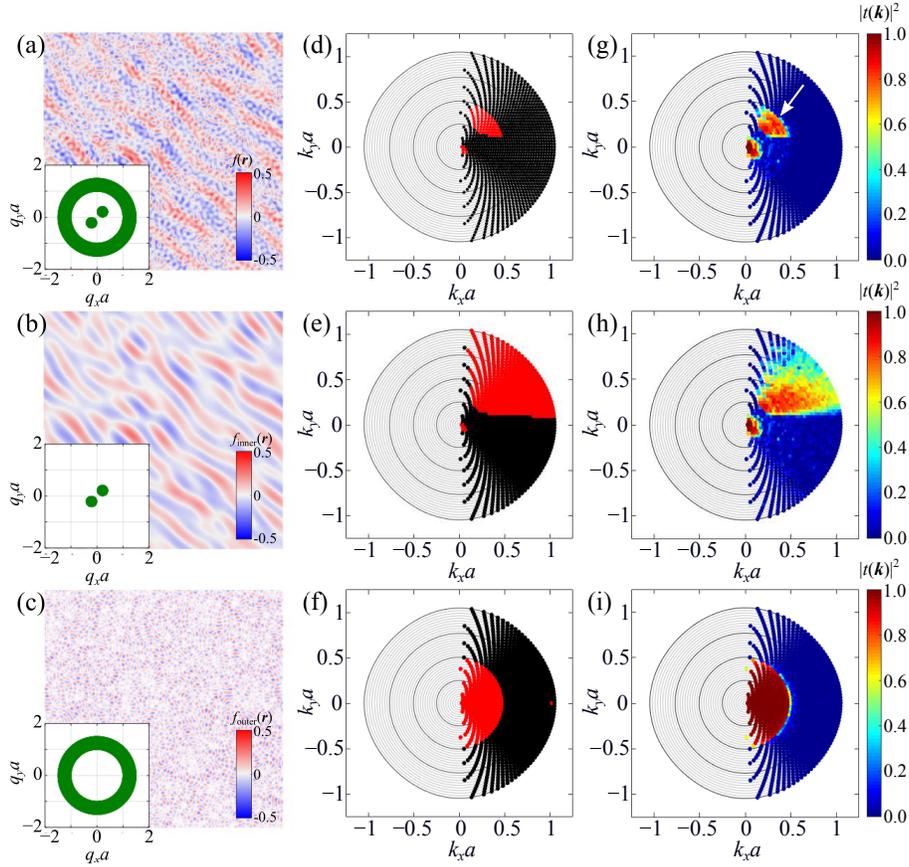}

\caption{Plot of the disorder function for (a) $f(\bm{r})$, (b) $f_{\text{inner}}(\bm{r})$,
and (c) $f_{\text{outer}}(\bm{r})$ with $f_{\text{inner}}$ rotated
by 45 degrees. The value of $f(\bm{r})$ is indicated by the color
bar in the bottom right corner. The bottom left corner insets show
the loci of the disorder modes in $\bm{q}$ space. The corresponding
distributions of the scattered (black dots) and unscattered (red dots)
$\bm{k}$ plane-wave states are shown in (d) to (f) while the $|t(\bm{k})|^{2}$
spectra for $L=20W$ are shown in (g) to (i) with the numerical values
represented by the color bar. The white arrow in (g) points to the
window of transparency for $f(\bm{r})$.}

\label{fig:45DegreeDisorderData}
\end{figure*}

\begin{figure*}
\includegraphics[width=12cm]{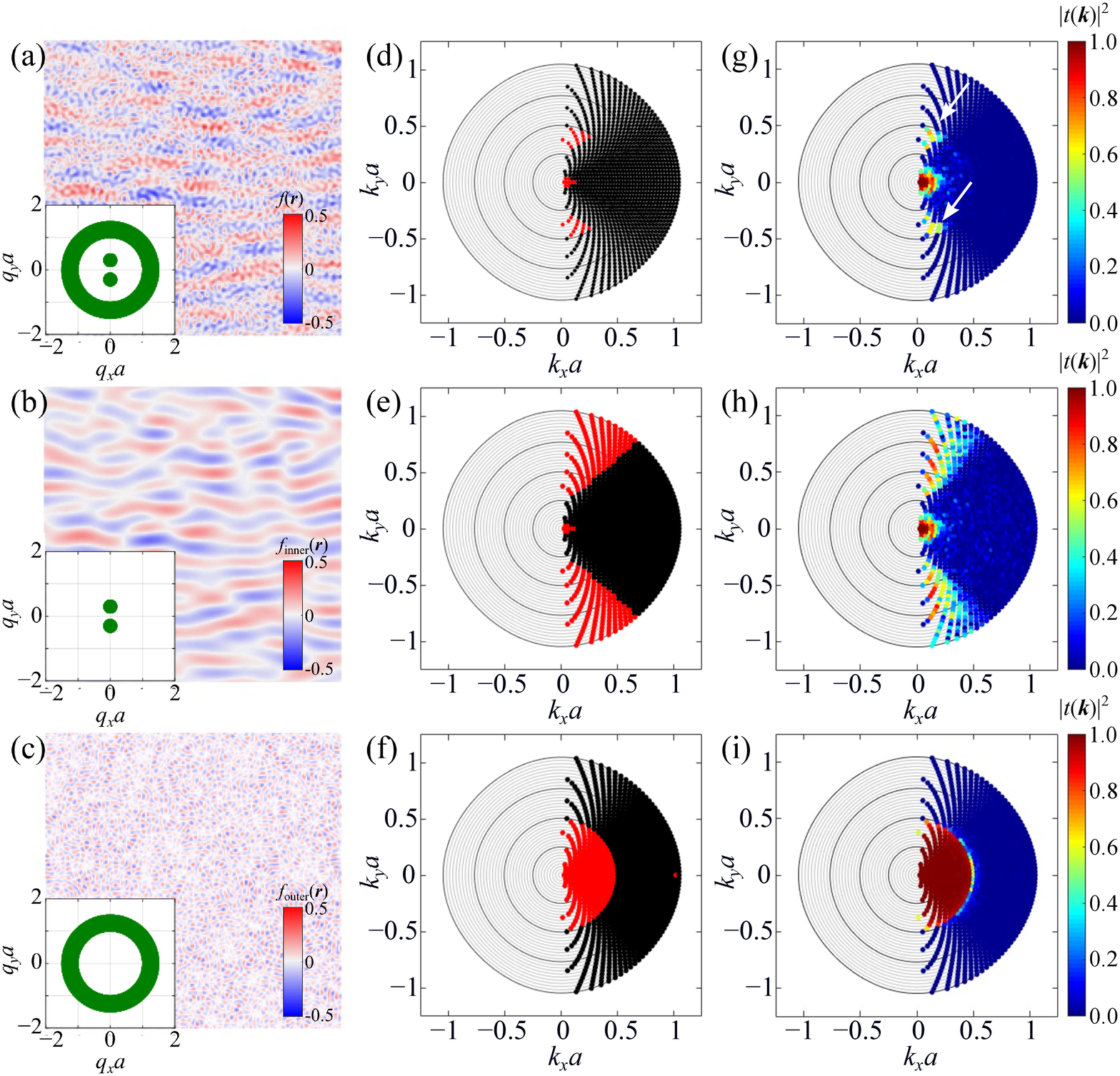}

\caption{Plot of the disorder function for (a) $f(\bm{r})$, (b) $f_{\text{inner}}(\bm{r})$,
and (c) $f_{\text{outer}}(\bm{r})$ with $f_{\text{inner}}$ rotated
by 90 degrees. The value of $f(\bm{r})$ is indicated by the color
bar in the bottom right corner. The bottom left corner insets show
the loci of the disorder modes in $\bm{q}$ space. The corresponding
distributions of the scattered (black dots) and unscattered (red dots)
$\bm{k}$ plane-wave states are shown in (d) to (f) while the $|t(\bm{k})|^{2}$
spectra for $L=20W$ are shown in (g) to (i) with the numerical values
represented by the color bar. The white arrows in (g) point to the
window of transparency for $f(\bm{r})$.}

\label{fig:90DegreeDisorderData}
\end{figure*}

We plot in Figs.~\ref{fig:45DegreeDisorderData} and \ref{fig:90DegreeDisorderData}
the simulation data for different $f_{\text{inner}}$. The data in
Fig.~\ref{fig:45DegreeDisorderData} are for an $f_{\text{inner}}$
that is like the one in Fig.~\ref{fig:CompositeDisorderTransmissionSpectra}
but rotated by 45 degrees with respect to the $x$ axis. The coherent
transmission spectra for $f$ and $f_{\text{inner}}$ are also rotated
by 45 degrees with respect to the $x$ axis. Hence, we do not observe
the window of transparency unless the angle of incidence is close
to 45 degrees. Similarly, the data in Fig.~\ref{fig:90DegreeDisorderData}
are for an $f_{\text{inner}}$ that is rotated by 90 degrees. Hence,
the window of transparency can only be observed at very oblique angles
of incidence close to 0 degree. 

\bibliography{PaperReferences,SuppInfo}

\end{document}